\documentclass{WileyChemistry-template}

\title{\raggedright Vibrationally {R}esolved {I}nner-{S}hell {P}hotoexcitation of the  {M}olecular {A}nion C$_\mathbf{2}^\mathbf{-}$}

 \author{
 \begin{minipage}{\textwidth}
S.~Schippers,\textsuperscript{*,[a]} P.-M.~Hillenbrand,\textsuperscript{+,[a]} A.~Perry-Sassmannshausen,\textsuperscript{+,[a]} T.~Buhr,\textsuperscript{+,[a]} S.~Fuchs,\textsuperscript{+,[a]} S.~Reinwardt,\textsuperscript{+,[b]} F.~Trinter,\textsuperscript{+,[c]} A.~M\"uller,\textsuperscript{+,[a]}  M.~Martins \textsuperscript{+,[b]}
\end{minipage}
}

\newcommand{\affiliation}{
 \begin{itemize}

 \item[{[a]}] Prof.\ Dr.\ S. Schippers, Dr.\ P.-M.~Hillenbrand, Dr.\ A.~Perry-Sassmannshausen, Dr.\ T.~Buhr, S. Fuchs, Prof.\ Dr.\ A.~M\"uller\\
     I.~Physikalisches Institut, Justus-Liebig-Universit\"{a}t Gie{\ss}en,  Heinrich-Buff-Ring 16, 35392 Giessen, Germany\\
     Email: stefan.schippers@physik.uni-giessen.de\\
     {http://www.uni-giessen.de/amp}

 \item[{[b]}]  S. Reinwardt, Dr.\ M. Martins\\
 Institut f\"{u}r Experimentalphysik, Universit\"{a}t Hamburg, Luruper Chaussee 149, 22761 Hamburg, Germany

 \item[{[c]}] Dr.\ F. Trinter\\
 Institut für Kernphysik, Goethe-Universit\"at Frankfurt am Main, Max-von-Laue-Strasse 1, 60438 Frankfurt am Main, Germany\\
 Molecular Physics, Fritz-Haber-Institut der Max-Planck-Gesellschaft, Faradayweg 4--6, 14195 Berlin, Germany

 \item[{[\texttt{+}]}] These authors contributed equally.
 \end{itemize}
}% end renewcommand

\newcommand{\keywords}{
	carbon dimer anion \textbullet\
	photon-ion merged-beams method \textbullet\
	inner-shell photoabsorption \textbullet\
	Franck-Condon analysis \textbullet\
    molecular potential curves
}% end newcommand

\renewcommand{\abstract}{
Carbon $1s$ core-hole excitation of the molecular anion C$_2^-$ has been experimentally studied at high resolution by employing the photon-ion merged-beams technique at a synchrotron light source.  The experimental cross section for photo--double-detachment shows a pronounced vibrational structure associated  with $1\sigma_u\to3\sigma_g$ and $1\sigma_g \to 1\pi_u$ core excitations of the C$_2^-$ ground level and first excited level, respectively. A detailed Franck-Condon analysis reveals a strong contraction of the C$_2^-$ molecular anion by 0.2~\AA\ upon this core photoexcitation. The associated change of the molecule's moment of inertia leads to a noticeable rotational broadening of the observed vibrational spectral features. This broadening is accounted for in the present analysis which provides the spectroscopic parameters of the C$_2^-$ $1\sigma_u^{-1}\,3\sigma_g^2\;{^2}\Sigma_u^+$  and $1\sigma_g^{-1}\,3\sigma_g^2\;{^2}\Sigma_g^+$ core-excited levels.
} %

\begin{document}

\twocolumn[\vspace{-1.5cm}\maketitle\vspace{-1cm}
	\textit{\dedication}\vspace{0.4cm}]
\small{\begin{shaded}
		\noindent\abstract
	\end{shaded}}
\section*{Introduction}

\begin{figure} [!b]
\begin{minipage}[t]{\columnwidth}{\rule{\columnwidth}{1pt}\footnotesize{\textsf{\affiliation}}}\end{minipage}
\end{figure}

With the advent of third-generation synchrotron light sources and x-ray free-electron lasers, inner-shell ionization of free molecules has become a topic of intense experimental research \cite{Ueda2003a,Hergenhahn2004,Feifel2011,Miron2011,Young2018,Ueda2019,Piancastelli2020}. Using \emph{neutral molecules} as targets, many fundamental questions have been addressed in recent years, e.g.,   the localization of the core hole \cite{Schoeffler2008},  complex many-electron relaxation effects \cite{Jahnke2007a,Havermeier2010,Ouchi2011,Trinter2013a},  or the role of the photon momentum in the molecular dissociation process \cite{Kukk2018,Kircher2019}. \emph{Molecular ions} have received much less attention despite their important role as transient species in many chemical environments such as flames \cite{Chen2019}, Earth's and Titan's ionospheres \cite{Shuman2015,Hsu2021}, or  interstellar gas clouds \cite{Snow2008,Millar2017}. This is because of the fact that only much lower target densities can be prepared for targets of free charged particles as compared to what can be achieved for neutral species. Nevertheless, recent progress in ion-beam and ion-trap techniques lead to first precision inner-shell studies with \emph{positively} charged molecular ions~\cite{Mosnier2016,Klumpp2018,Bari2019,Schubert2019,Kennedy2019,Lindblad2020,Couto2020,Carniato2020,Martins2021,Lindblad2022,Schwarz2022}.

Here, we report on  $1s$ inner-shell photoexcitation of the  \emph{negatively} charged molecular ion C$_2^-$.  Much experimental work on  C$_2^-$ has addressed the valence shell using a number of  spectroscopic techniques\cite{Leutwyler1982,Hefter1983,Mead1985,Rehfuss1988,Ervin1991,Pedersen1998a,Bragg2003,Iida2020,Iizawa2022}.  Recently, C$_2^-$ has been identified as a promising molecular species for laser cooling  \cite{Yzombard2015}, and the corresponding transitions have been studied with high precision \cite{Noetzold2022}. In the present study, we have measured relative cross sections for $1s$ inner-shell photo--double-detachment (PDD) of C$_2^-$ anions, i.e., for the process
\begin{equation}
h\nu + \textrm{C}_2^- \to \textrm{C}_2^+ + 2e^-.
\end{equation}
We used a sufficiently high photon-energy resolving power that allowed us to resolve transitions between individual vibrational levels, thus providing detailed insight into the highly correlated nuclear and electronic relaxation dynamics,  that sets in after the initial creation of a C~$1s$ core hole.  Previous work on inner-shell photoabsorption by molecular anions was either confined to inner-valence shells \cite{Bilodeau2006,Laws2019} or, for photoabsorption by K-shell electrons, used rather low photon-energy resolving powers \cite{Berrah2011,Douguet2020}.

\section*{Experiment{al Setup}}

The present experiment was performed at the  photon-ion merged-beams setup PIPE \cite{Schippers2014,Mueller2017,Schippers2020}, a permanently installed end-station at the photon beamline P04 \cite{Viefhaus2013} of the PETRA\,III synchrotron light source, operated by DESY in Hamburg, Germany. Using the same procedures as in a previous experiment with atomic C$^-$ anions \cite{Perry-Sassmannshausen2020}, a C$_2^-$ ion beam was generated with a Cs-sputter ion source, accelerated to  a kinetic energy of 6~keV, and magnetically analyzed  for isolating ions with the desired mass-over-charge ratio. Subsequently, the mass-over-charge-selected $^{12}$C$^{12}$C$^-$ ion beam was collimated and electrostatically deflected such that it moved coaxially with the counter-propagating photons over a distance of  $\sim$1.7~m length. The collimated C$_2^-$ ion current in the photon-ion interaction region amounted to typically 10 nA,  and the photon flux was $9\times10^{11}$~s$^{-1}$  at a photon energy of 280~eV and a photon-energy spread of  500~meV. This rather low photon flux at the photon energies of present interest resulted from strong photoabsorption by carbon contaminations on the surfaces of the  photon beamline's optical components.

The photon-energy scale was calibrated by performing photoabsorption measurements in N$_2$ and Ne gases and by linearly scaling the photon energy such that the positions of the measured absorption resonance features  matched known energies  \cite{Sodhi1984a,Mueller2017,Mueller2018c}. An additional energy correction accounted for the Doppler shift associated with the directed  movement of the ions in the primary C$_2^-$ ion beam relative to the photon beam. The resulting uncertainty of the present calibrated energy scale amounts to $\pm0.2$~eV.

The elongated photon-ion interaction volume ensured that the number of photoionization events was sufficiently large for acquiring an acceptably low level of statistical uncertainty in a reasonable amount of time despite the diluteness of the ionic target.  Behind the photon-ion interaction region, the C$_2^+$ reaction products were magnetically separated from the C$_2^-$ primary ion beam and directed onto a single-particle detector. Background resulting from charge-changing collisions with residual-gas  particles was determined by separate measurements in absence of the photon beam.  {At the maximum of the first resonance in Figure~\ref{fig:C2}, the signal count rates amounted  to about 1~kHz and  150~Hz at photon-energy spreads of 500~meV and 50~meV, respectively. The background count rate was about 10~Hz in both cases.} The background-subtracted photo-product count rate was normalized to the primary ion current, which was continuously measured with a Faraday cup, and to the photon flux, that was monitored by a calibrated photodiode. This procedure resulted in the relative cross sections for PDD of  C$_2^-$ anions that are displayed in Figure~\ref{fig:C2}.

\section*{Results and Discussion}

\begin{figure}
\includegraphics[width=\columnwidth]{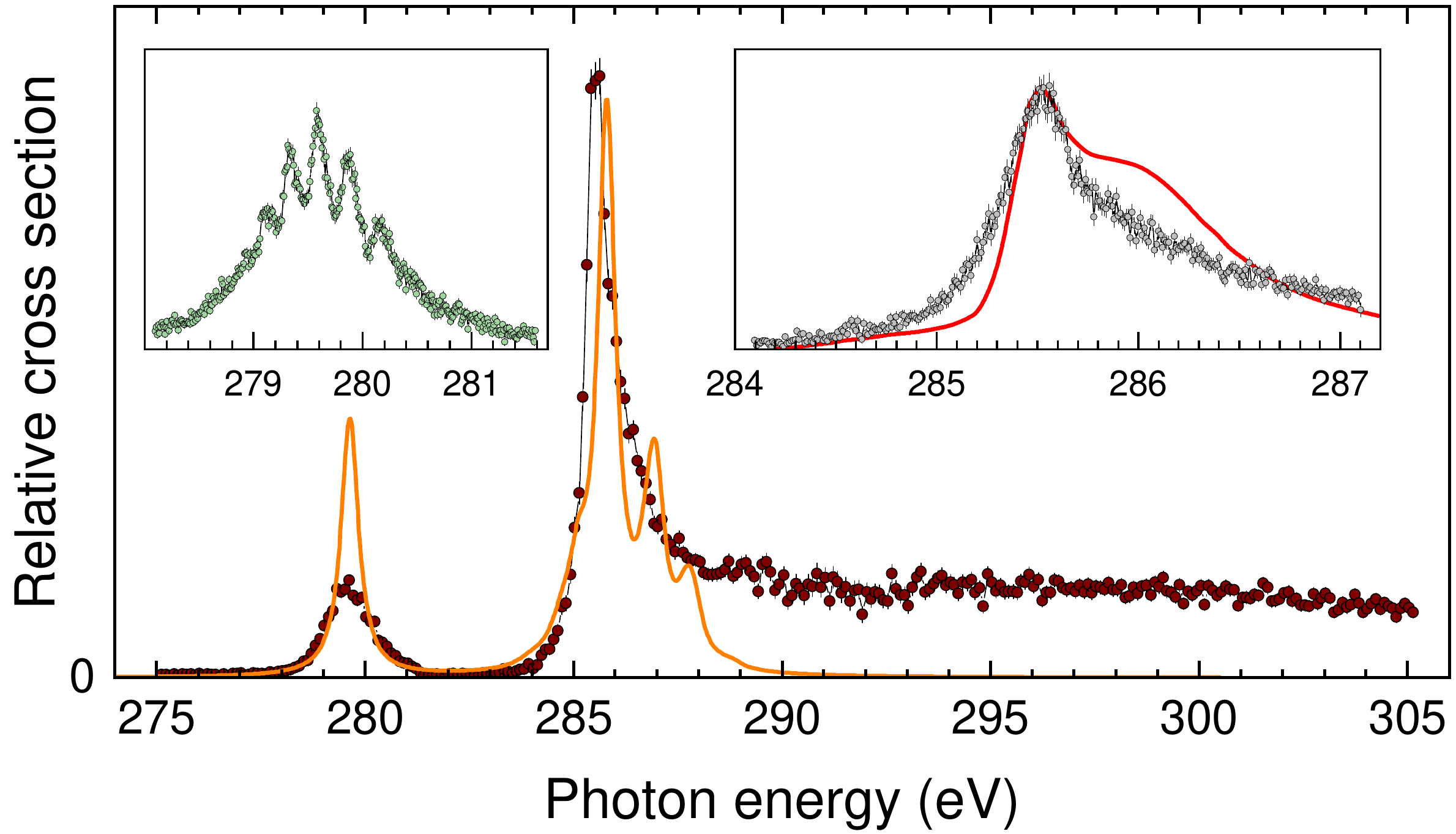}
\caption{\label{fig:C2}Experimental cross sections (symbols) for PDD of C$_2^-$. The main panel provides an overview over the entire experimental photon-energy range which was scanned at a photon-energy spread of $\Delta E_\mathrm{exp}\approx 500$~meV. The orange full line is  the  theoretical absorption cross section calculated within this work (see text) scaled and shifted in energy (by +10.5~eV) to match the experiment. The insets display high-resolution measurements ($\Delta E_\mathrm{exp}\approx 50$~meV) of the  resonance features. In all cases, the statistical error bars represent one standard deviation. All cross section scales are linear. The red full line in the right inset is the theoretical cross section for photodetachment of C$_2^-$ by Douguet et al.~\cite{Douguet2020}, scaled to the experimental cross section and shifted by $-0.1$~eV.}
\end{figure}

The main panel of Figure~\ref{fig:C2} provides an overview over the PDD cross section in the vicinity of the C$_2^-$ $1\sigma$  ionization threshold which has been theoretically predicted at  284.1~eV \cite{Douguet2020}.  The cross section exhibits two resonance features, one $\sim$4.5~eV below this threshold and one $\sim$1.4~eV above. These two features were already observed in the experimental data  communicated by Berrah and Bilodeau to Douguet et al.~\cite{Douguet2020}. These data were measured at  a similar  photon-energy spread as in the present overview scan.  To obtain more accurate information about the resonance features, we have measured both features also at a lower  photon-energy spread of approximately 50~meV. This reveals the vibrational structure of the first resonance (left inset of Figure~\ref{fig:C2}). However, no such vibrational structure is discernible for  the second resonance (right inset of Figure~\ref{fig:C2}).

The main panel of Figure~\ref{fig:C2} also displays a theoretical cross section for photoabsorption of C$_2^-$. It was calculated with the ORCA quantum chemistry program package\cite{Neese2020} (version 5.0.3)  using the def2-TZVPP\cite{Weigend2005} and def2/J\cite{Weigend2006} basis sets as well as 20 uncontracted Gaussian $s$- and $p$-type functions \cite{Kaufmann1989} for a better description of the Rydberg states. In the TDDFT calculations, the CAM-B3LYP\cite{Yanai2004} functional was employed together with the RIJCOSX\cite{Neese2009} approximation for the evaluation of matrix elements. For the comparison with the experimental data the calculated cross section was convolved with a Gaussian with a full width at half maximum (FWHM) of 0.5~eV, multiplied with a constant factor, and shifted by 10.5~eV towards higher energies.

Before discussing the comparison between experiment and theory, it should be recalled that the measured double de\-tach\-ment cross section does not directly correspond to the calculated absorption cross section. Next to double de\-tach\-ment, other processes do contribute to photoabsorption as well such as single detachment or molecular breakup into neutral or charged atomic fragments. Generally, the branching ratios for the various final channels depend on the photon energy. For example, above the threshold for direct C $1s$ detachment, net single detachment becomes improbable since the C $K$-shell hole that is formed by the primary detachment process will be most probably filled by an Auger process leading either to double detachment where the molecule is left intact (this is what has been observed in the present experiment) or to fragmentation.

As for the present ORCA calculation, there is agreement between experiment and theory concerning the energy separation between the two experimentally observed resonance features (Figure~\ref{fig:C2}). The calculated relative strengths of these two features agree less with the experimental findings. This may be attributed to different double-detachment branching ratios for both resonance features. Moreover, the ion beam probably contained a sizeable fraction of metasta\-ble C$_2^-$ anions (see below) whereas, in the calculation, it was assumed that all ions were in their ground level. The asymmetric shape of the above-threshold resonance is explained by the presence of multiple resonances of decreasing strength with increasing energy. However, in the experimental cross section, this resonance structure is more smeared out than predicted by theory. The theoretical calculation does not account for direct ionization of a C $1s$ electron which dominates the experimental cross section above about 288~eV.

The cross section for C$_2^-$ photodetachment has been calculated previously by Douguet et al.~\cite{Douguet2020} who considered only the strong above-threshold resonance. The red full line in the right inset of Figure~\ref{fig:C2} displays their result, scaled to the present experimental cross section and shifted by -0.1~eV on the photon-energy axis. This shift is within the present $\pm0.2$~eV experimental uncertainty. While experiment and theory agree about the position of this resonance feature, there are slight differences concerning its shape, in particular, on the high-energy side of the peak where the theoretical calculation predicts a pronounced shoulder, which is not present in the experimental data. However, one again has to take note of the fact  that  different final channels were considered in theory (single detachment) and experiment (double detachment). According to the calculations, the resonance classifies as a shape resonance associated with $1\sigma_g\to k\pi_u$ dipole excitations to short-lived levels of $^2\Pi_u$ symmetry. These are weakly bound in a shallow potential supporting only a few narrowly spaced vibrational levels, which cannot be resolved even by our high-resolution measurement.

The lowest $1\sigma$ core excitation of the \linebreak C$_2^-$  $1\sigma_g^2\,1\sigma_u^2\,2\sigma_g^2 \,2\sigma_u^2\,1\pi_u^4\,3\sigma_g\;X\,^2\Sigma_g^+$ ground level is a $1\sigma_u\to3\sigma_g$ excitation to the  $1\sigma_u^{-1}\,3\sigma_g^2\;^2\Sigma^+_u$ level. Corresponding vibrationally resolved photoabsorption resonances were observed for isoelectronic N$_2^+$ \cite{Lindblad2020}. Therefore, we assign the vibrationally resolved C$_2^-$ resonance structure to the same electronic transition, despite of the fact that the vibrational structure of the positive ion differs significantly from the present one for the negatively charged carbon dimer. For N$_2^+$, the strongest vibrational transition is associated with the $v'=0$ excited vibrational level and the contributions by higher vibrational levels decrease monotonically with increasing vibrational quantum number $v'$ such that individual peaks can only be discerned for $v'=0$, $v'=1$, and $v'=2$. For C$_2^-$, the vibrational distribution is much broader attaining its maximum at $v'=5$ (Figure~\ref{fig:C2fit1}). This behavior is related to a substantial change of the C$_2^-$ bond length upon $1\sigma_u \to 3\sigma_g$ core excitation as revealed by the Franck-Condon analysis that is discussed in the following.

\subsection*{Franck-Condon {A}nalysis}

Our Franck-Condon analysis consists of fitting a sum of Voigt line profiles $V(E,\Delta E_\mathrm{exp},\gamma,E_{vv'},f_{vv'})$ to our experimental high-resolution data with each profile representing a transition between a vibrational level $v$ of the initial electronic level and a vibrational level $v'$ of the core-excited potential curve. The Voigt profiles are functions of the photon energy $E$. Their widths are characterized by the experimental photon-energy spread $\Delta E_\mathrm{exp}=0.05$~eV (Gaussian full width at half maximum) and by the Lorentzian width $\gamma$ that is associated with the core-hole lifetime of the core-excited level. In principle, the Lorentzian width should also depend on $v$ and $v'$. However, the present data do not suggest that there is a noticeable dependence of $\gamma$ on the vibrational quantum numbers.  Therefore, it is not taken into account in our present analysis. The individual Voigt profiles are centered at the vibrational transition energies  $E_{vv'}$. The relative strengths of the transitions are given by the corresponding Franck-Condon factors  $f_{vv'}$ which are calculated from the potential parameters of the lower and upper potential curves (Morse potentials).

Concretely, the fit function in the present Franck-Condon analysis was
\begin{equation}\label{eq:fit1}
\sigma(E) = b_0+b_1E +\sum_{k=1}^{k_\mathrm{max}} S_k \sigma_k(E)
\end{equation}
where the coefficients $b_0$ and $b_1$ account for the {continuum cross section for $L$-shell detachment} and the sum extends over different electronic transitions enumerated by the summation index $k$ and the factors $S_k$ denote the associated apparent (relative) transition strengths. Each individual electronic transition $k$ contributes the absorption cross section
\begin{eqnarray}
\sigma_k(E)&=& \sum_{v=0}^{v_\mathrm{max}}n_{kv}\sum_{v'=0}^{v'_\mathrm{max}}\sum_{J=0}^{J_\mathrm{max}}\sum_{J'=J-1}^{J+1}s_{kJJ'} \times \label{eq:fit2}\\ & & V\left(E,\Delta E_\mathrm{exp},\gamma_k,E_{kvv'}+\Delta E_{kJJ'},f_{kvv'}\right),\nonumber
\end{eqnarray}
with $n_{kv}$ denoting the fractional populations of the initial vibrational level $v$ of the electronic transition $k$.  The sum over the rotational quantum numbers $J$ and $J'$ and the quantities $s_{kJJ'}$ and $\Delta E_ {kJJ'}$ account for rotational broadening as detailed below. The calculation of the Franck-Condon factors between two displaced Morse potentials follows the prescription of López et al.~\cite{Lopez2002}. It involves a numerical integration which is carried out by  using an adaptive Gauss-Kronrod algorithm and extended precision arithmetic as implemented in the boost C$^{++}$ libraries \cite{boost}. We verified the accuracy of our numerical integration procedure by reproducing the Franck-Condon factors tabulated by López et al.~\cite{Lopez2002}.

{In the fits below, the parameters $b_0$, $b_1$, $S_k$ from Equation~\ref{eq:fit1}, $\gamma_k$ from Equation~\ref{eq:fit2}, as well as the parameters $\hbar\omega_e$, $\hbar\omega_e\chi_e$, and $R_e$ of the core-excited Morse potential curves \cite{Lopez2002} were varied simultaneously. In each fit step, the momentary values of the Morse parameters were used for the calculation of the vibrational energies $E_{kvv'}$ and the Franck-Condon factors $f_{kvv'}$ appearing  in Equation~\ref{eq:fit2}. An additional free fit parameter was the temperature pertaining to the vibrational and rotational degrees of freedom. This temperature determined the values of $n_{kv}$ and $s_{kJJ'}$ from Equation~\ref{eq:fit2} as explained below.}

\subsection*{Fit {C}onsidering {O}ne {E}lectronic {T}ransition}

\begin{figure}[b!]
\includegraphics[width=\columnwidth]{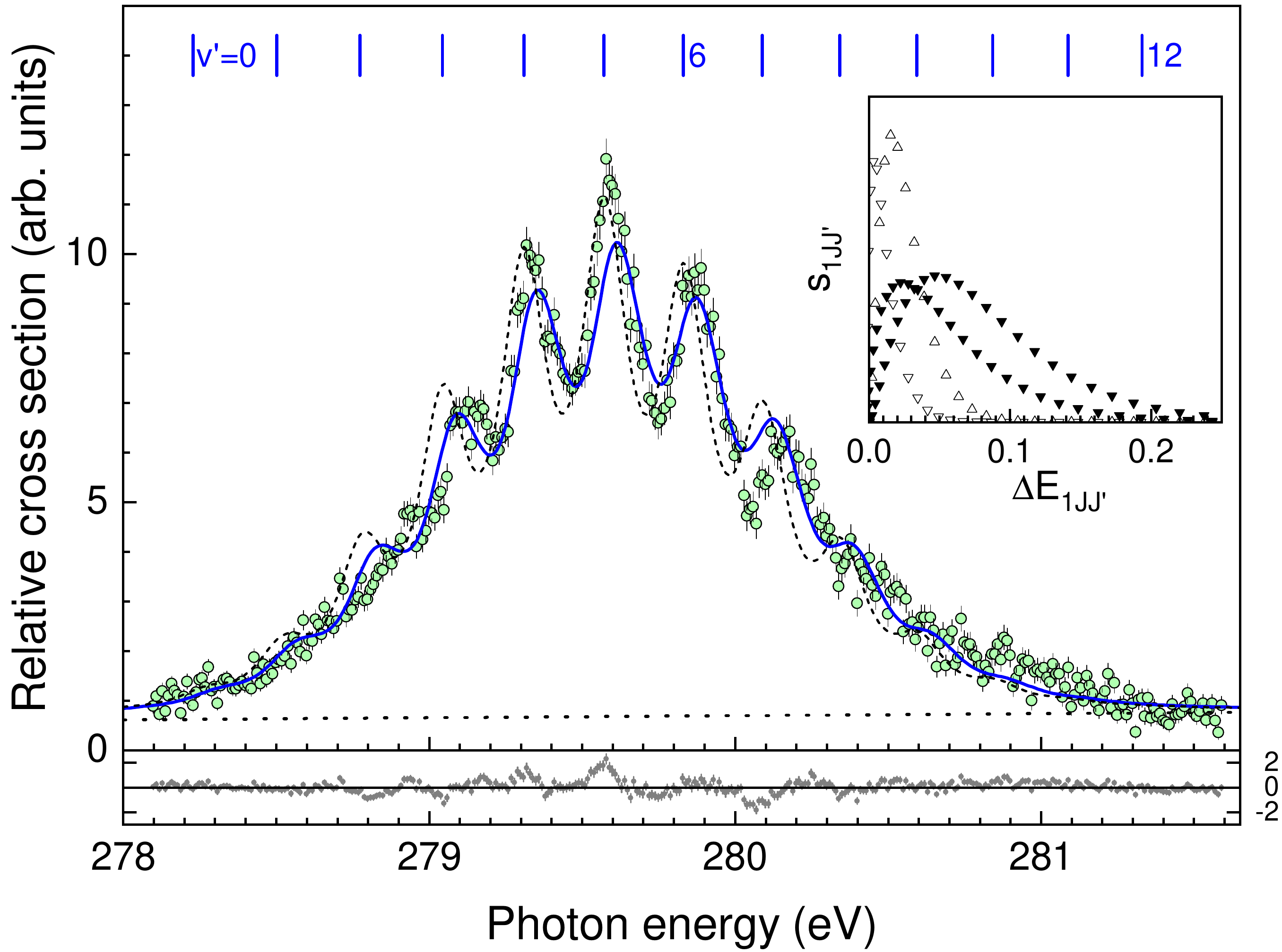}
\caption{\label{fig:C2fit1}Experimental cross section for PDD of C$_2^-$ (symbols with statistical error bars representing one standard deviation) and the fit (blue full curve) of  Equations~\ref{eq:fit1} and \ref{eq:fit2} with $k_\mathrm{max}=1$, $v_\mathrm{max}=1$ and $v'_\mathrm{max}=12$ for only the $1\sigma_u \to 3\sigma_g$ core excitation of the $X\,^2\Sigma_g^+$ ground level. The dotted line {represents the continuum cross section for $L$-shell detachment as obtained from the fit}.  The vertical bars mark the vibrational energies for $v=0 \to v'$ transitions that result from the fit. The dashed curve is obtained if the rotational temperature is set to zero and all other fit parameters remain unchanged. It allows one to assess the influence of rotational broadening on the shape of the spectrum. The inset shows the underlying rotational energy (Equation~\ref{eq:Erot}) distributions  for the $J\to J'=J+1$ (triangles up) and $J\to J'=J-1$ (triangles down) transitions with even $J$ between the lower $X\,^2\Sigma_g^+$ and the upper $^2\Sigma_u^+$ C$_2^-$ rotational levels for the temperatures $T=1100$~K (full symbols) as used in the fit and $T=300$~K for comparison (open symbols). Odd $J$ values and $J\to J'=J$ transitions do not occur because of nuclear-spin statistics. The lower panel displays the residuals of the fit together with the experimental error bars.  The reduced $\chi^2$ of the fit is 3.19.  }
\end{figure}

In a first fit, only the $1\sigma_u \to 3\sigma_g$ transition from the $X\,^2\Sigma^+_g$ electronic ground level was considered. The potential parameters of this level were taken from recent theoretical work \cite{Shi2016}, where excellent agreement with the available experimental data  \cite{Mead1985,Rehfuss1988} was achieved. The potential parameters of the core-excited potential curve were varied in the fit. This fit result is displayed in Figure~\ref{fig:C2fit1}. It reproduces the overall resonance structure, albeit not in every detail. Nevertheless, the fit reveals that the maximum vibrational transition strength occurs for the $v=0 \to v'=5$ transition and that this is related to a strong decrease of the C$_2^-$ bond length by almost~$-0.2$~\AA\ upon $1\sigma_u \to 3\sigma_g$ core excitation. For isoelectronic N$_2^+$, the corresponding decrease was found to be only $-0.04$~\AA\ \cite{Lindblad2020}, leading to a drastically different vibrational resonance structure as compared to C$_2^-$.

The large change of the C$_2^-$ bond length upon inner-shell excitation entails a large change of the  molecule's moment of inertia. This and the high molecular temperature of $\sim$1100~K (see below) leads to a noticeable rotational broadening of the vibrational spectral structures. This effect was not considered in the analysis of the vibrationally resolved inner-shell detachment of N$_2^+$ \cite{Lindblad2020}, where the photo-induced change of bond length was found to be comparatively moderate and where the molecular ions were internally colder. In our fit, we have quantified the rotational broadening within the rigid-rotator approximation. Accordingly, the rotational constant $B_e$ scales with $R_e^{-2}$.  With $\Delta R_e = -0.2$~\AA, the rotational constant $B_e'= B_e[R_e/(R_e+\Delta R_e)]^2$ of the upper $^2\Sigma_u^+$ level is larger than $B_e$ of the $X\,^2\Sigma_g^+$  ground level by more than 40\%. The associated  change of the C$_2^-$ rotational energy upon $1s$ core excitation depends on the rotational quantum numbers $J$ and $J'$ of the lower and upper levels, respectively. It amounts to
\begin{equation}\label{eq:Erot}
\Delta E_{JJ'}=hc\left[B_e'J'(J'+1)-B_eJ(J+1)\right].
\end{equation}
Assuming a Boltzmann distribution for the rotational levels of the $X\,^2\Sigma_g$ electronic ground level and accounting for nuclear-spin statistics \cite{Hertel2015} as well as for the relative rotational transition strengths (Hönl-London factors \cite{Hansson2005} denoted as $s_{kJJ'}$ in Equation~\ref{eq:fit2}), one arrives at the rotational-energy distributions that are shown in the inset of Figure~\ref{fig:C2fit1}. The distribution for a temperature of 1100~K is strongly asymmetric. The mean energy shift amounts to 54.2~meV and its standard deviation to 3.2~meV, thus, leading to a significant shift and broadening of the vibrational resonances as can be seen from a comparison of the fit curve with the dashed curve in Figure~\ref{fig:C2fit1} which was calculated from the same set of fit parameters but does not account for rotational effects.

\subsection*{Fit {C}onsidering {T}wo {E}lectronic {T}ransitions}

\begin{figure}
\includegraphics[width=\columnwidth]{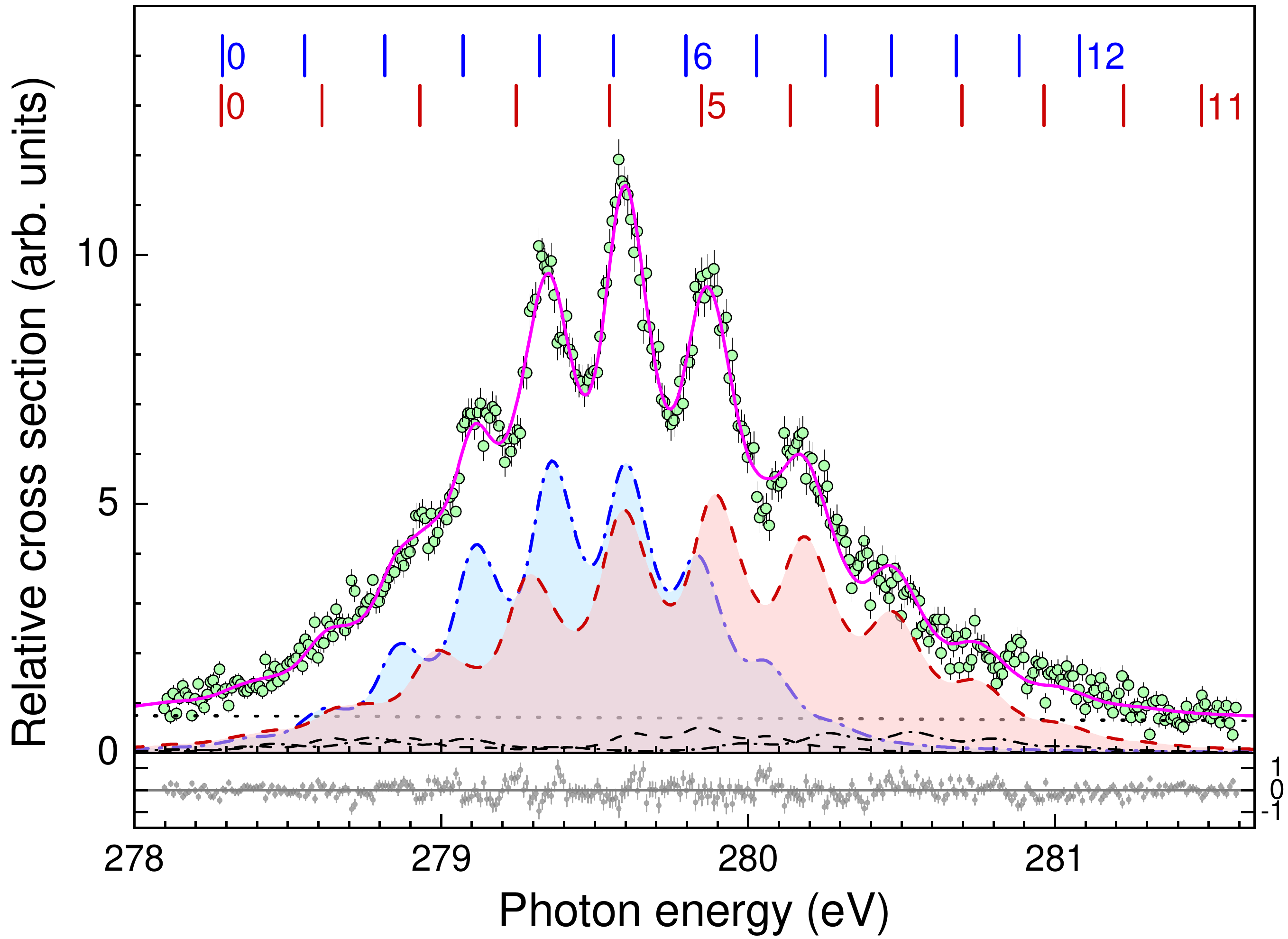}
\caption{\label{fig:C2fit2}Experimental cross section for PDD of C$_2^-$ (symbols with statistical error bars representing one standard deviation) and the fit (pink full curve)  of  Equations~\ref{eq:fit1} and \ref{eq:fit2} with $v_\mathrm{max}=1$ and $v'_\mathrm{max}=12$ {for two electronic transitions ($k_\mathrm{max}=2$). The blue dashed-dotted and red dashed shaded curves represent $S_k\sigma_k(E)$ for $k=1$ and $k=2$, respectively}. The dash-dotted and dashed thin black lines are the respective contributions by the $v=1$ initial vibrational level. The dotted line {represents the continuum cross section for $L$-shell detachment as obtained from the fit.} The vertical bars mark the vibrational energies $E_{k0v'}$  with the upper (lower) row corresponding to  $k=1$ ($k=2$). They are labelled by the vibrational quantum number $v'$ of the excited electronic level. The lower panel displays the residuals of the fit together with the experimental error bars. The reduced $\chi^2$ of the fit is 1.80.}
\end{figure}

\begin{figure}[b!]
\includegraphics[width=\columnwidth]{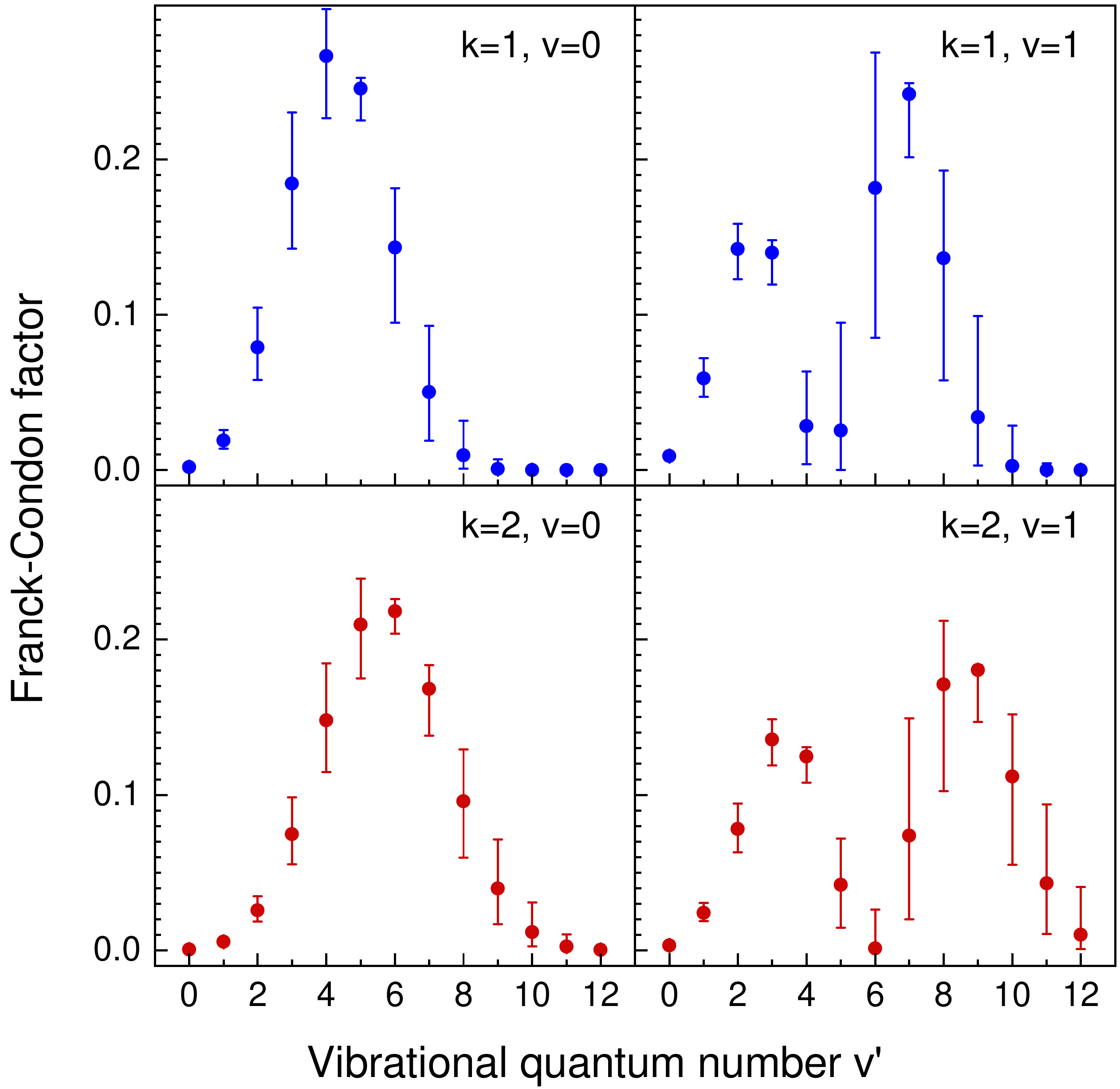}
\caption{\label{fig:FC}Franck-Condon factors $f_{kvv'}$ resulting from the fit of Equations~\ref{eq:fit1} and \ref{eq:fit2} to the experimental data (Figure~\ref{fig:C2fit2}). The vibrational quantum numbers $v$ and $v'$ refer to the lower and upper levels, respectively,  of the electronic transitions $k$. The error bars were obtained from varying uncorrelatedly the parameters $R_e$ ,  $\hbar\omega_e$, and $\hbar\omega_e\chi_e$ of the upper levels within the respective uncertainty intervals from Table~\ref{tab:fit}.}
\end{figure}

\begin{table*}
\caption{\label{tab:fit} Spectroscopic parameters resulting from the fit of Equations~\ref{eq:fit1} and \ref{eq:fit2} to the experimental data displayed in Figure~\ref{fig:C2fit2}, i.e.,  the minimum $T_e$ of the potential curve,   the harmonic frequency $\omega_e$, the anharmonicity coefficient  $\omega_e\chi_e$, the equilibrium bond length $R_e$, and the natural line width $\gamma$. The depths and the inverse ranges of the Morse potentials were calculated as $D=\hbar\omega_e^2/(4 \omega_e\chi_e)$ and $\beta=\sqrt{2\mu\omega_e\chi_e/\hbar}$  \cite{Lopez2002}. In the latter expression,  $\mu$ denotes the reduced mass. Also given are the rotational constants $B_e$ and the differences $\Delta R_e$ between the bond lengths of the lower and upper levels. The parameters for the $X\,^2\Sigma^+_g$ ground level and the $A\;^2\Pi_u$ metastable level were taken from~\cite{Shi2016}. Numbers in parentheses signify the (uncorrelated) errors from the fit. In addition, $T_e$ is subject to the systematic experimental uncertainty of the energy calibration of $\pm$0.2 eV.}
{\scriptsize \begin{tabular}{cclllllllcl}
\toprule
 $k$  & Level &   \multicolumn{1}{c}{$T_e$ [eV]} &    \multicolumn{1}{c}{$\hbar\omega_e$ [meV]} & \multicolumn{1}{c}{$\hbar\omega_e\chi_e$ [meV]}  & \multicolumn{1}{c}{$D$ [eV]} & \multicolumn{1}{c}{$\beta$ [\AA$^{-1}$]}  &  \multicolumn{1}{c}{$B_e$ [cm$^{-1}$]} & \multicolumn{1}{c}{$R_e$ [\AA]} & \multicolumn{1}{c}{$\Delta R_e$ [\AA]} &  \multicolumn{1}{c}{$\gamma$ [meV]} \\
\midrule
  1  & $(1\pi_u^4\,3\sigma_g)\,X\,^2\Sigma^+_g$                &    $\phantom{27}0.0$      &  ~~~$220.84$ & ~~~$1.4289$  &  $8.5330$  & ~$2.0253$ & ~$1.7438$  & $1.2689$   & $-$ & -\\
  1  & $1\sigma_u^{-1}\,1\pi_u^4\,3\sigma_g^2\;{^2}\Sigma_u^+$ &  $278.26(2)$  &  ~~~$274(8)$ & ~~~$3.2(6)$  &  $5.9(12)$ & ~$3.0(3)$ & ~$2.42(1)$ & $1.077(2)$ & $-0.191(2)$ & $124(12)$ \\[1ex]
  2  & $(1\pi_u^3\,3\sigma_g^2)\,A \;^2\Pi_u$                  &    $\phantom{27}0.49655$  &  ~~~$206.90$ & ~~~$1.3387$  &  $7.9939$  & ~$1.9604$ & ~$1.6424$  & $1.3075$   & $-$ & -\\
  2  & $1\sigma_g^{-1}\,1\pi_u^4\,3\sigma_g^2\;{^2}\Sigma_g^+$ &  $278.38(3)$  &  ~~~$342(6)$ & ~~~$3.7(5)$  &  $7.9(14)$ & ~$3.3(3)$ & ~$2.30(2)$ & $1.105(4)$ & $-0.202(4)$ & $154(21)$ \\
\bottomrule
\end{tabular}
}% end small
\end{table*}

\begin{figure}
\centering{\includegraphics[width=0.9\columnwidth]{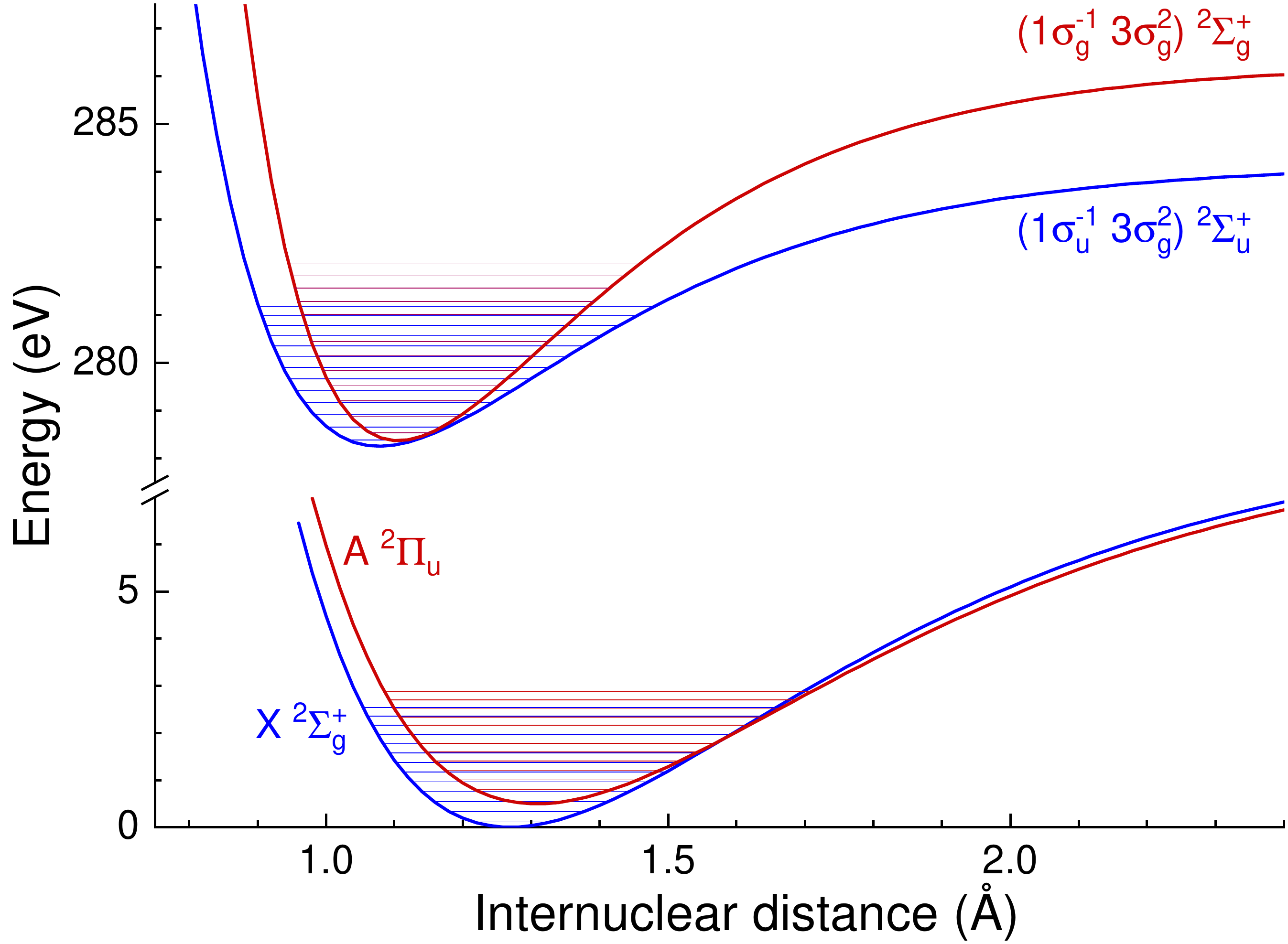}}
\caption{\label{fig:C2pot}Morse potentials with their 12 lowest vibrational levels resulting from the spectroscopic parameters given in Table~\ref{tab:fit}. }
\end{figure}

The agreement between fitted and experimental curve in Figure~\ref{fig:C2fit1} is not satisfying. An attempt to include nonthermal populations of higher initial vibrational levels resulted in a smearing out of the vibrational resonance structures and was not pursued any further. Instead, we considered a second electronic transition starting from the first electronically excited level  $1\sigma_g^2 1\sigma_u^2 2\sigma_g^2 2\sigma_u^2 1\pi_u^3 3\sigma_g^2\;A\,^2\Pi_u$. In the Cs-sputter source, C$_2^-$ anions are formed, when the carbon sputtered from the cathode traverse the Cs monolayer covering the cathode. The electron transfer of Cs valence electrons to the C$_2$ dimers generates C$_2^-$ anions in the electronic ground state as well as in electronically excited states \cite{Pedersen1998a}. The $A\,^2\Pi_u$ level has  an excitation energy of $\sim$0.5~eV and lifetimes of 50~$\mu$s for $v=0$ and 40~$\mu$s for $v=1$ \cite{Rosmus1984,Shi2016}. These lifetimes are of the order of the flight time of the ions from the ion source to the merged-beams interaction region. Therefore, it must be expected that a fraction of the C$_2^-$ anions is in this metastable electronically excited level. The second electronically excited level of C$_2^-$, the B$\,^2\Sigma_u^+$ level with an excitation energy of $\approx2.3$~eV, has a lifetime of less than 80~ns \cite{Leutwyler1982,Rosmus1984,Shi2016} which is much shorter than the ions' flight time. Therefore, this level was not taken into account in the present analysis.

The lowest core-excitation channel of the $A\,^2\Pi_u$ level is the $1\sigma_g \to 1\pi_u$ transition to the $1\sigma_g^{-1} 1\sigma_u^2 2\sigma_g^2 2\sigma_u^2 1\pi_u^4 3\sigma_g^2\;^2\Sigma_g^+$ level. Inclusion of both the  $1\sigma_u \to 3\sigma_g$ and $1\sigma_g \to 1\pi_u$ transitions from the $X\,^2\Sigma_g^+$ and $A\,^2\Pi_u$ levels to the nearly degenerate $1\sigma_u^{-1}\,3\sigma_g^2\;^2\Sigma_u^+$ and $1\sigma_g^{-1}\,3\sigma_g^2\;^2\Sigma_g^+$ levels, respectively, resulted in the much improved fit displayed in Figure~\ref{fig:C2fit2}. The potential parameters that were determined by  this fit are provided in Table~\ref{tab:fit} and the respective Frank-Condon factors $f_{kvv'}$ (Equation~\ref{eq:fit2}) are plotted in Figure~\ref{fig:FC}.

The fit suggests that the contribution of the $A\,^2\Pi_u$ level to the PDD cross section is substantial ($55\pm7\%$). It should be kept in mind that this percentage does not correspond to the initial population of this level. It also reflects the relative line strength of the core-exciting $1\sigma_g \to 1\pi_u$ transition which might be larger than the one of the $1\sigma_u \to 3\sigma_g$ transition from the $X\,^2\Sigma_g^+$ ground level.

Figure~\ref{fig:C2pot} displays the Morse potentials that correspond to the $D$ and $\beta$ values from  Table~\ref{tab:fit}. The contraction of the molecule upon $1\sigma$ core excitation is obvious. The minima of the core-excited potential curves are shifted by $-0.19$ and $-0.20$~\AA , respectively, to considerably lower internuclear distances as compared to the $X\,^2\Sigma_g^+$ and $A\,^2\Pi_u$ potentials. These changes in bond length are considerably larger than what has been reported for isoelectronic N$_2^+$, where the respective value for the core excitation of the $X\,^2\Sigma_g^+$ ground level is $-0.04$~\AA\ \cite{Lindblad2020}. The difference of 0.15~\AA\ between $\Delta R_e$ for the core excitation of the C$_2^-$ and N$_2^+$ $X\,^2\Sigma_g^+$ levels stems exclusively from the difference in the $X\,^2\Sigma_g^+$ equilibrium bond lengths which are 1.27 and 1.12~\AA, respectively. Notably, the bond length of 1.08~\AA\ of the core-excited $^2\Sigma_u^+$ level is the same for both molecular species, i.e., upon core photoexcitation, C$_2^-$ seems to loose its anionic character.

It is somewhat surprising that the potential curves of the two core-excited $1\sigma_{u/g}^{-1}\, 3\sigma_g^2\;^2\Sigma_{u/g}^+$ levels should be different since one would expect a near degeneracy. It should, however, be noted that the differences between the Morse parameters $D$ and $\beta$ for both levels are within their mutual fit uncertainties (Table~\ref{tab:fit}). We also tried a fit where we imposed the additional constraint that $\hbar\omega_e$, $\hbar\omega_e\chi_e$, and $R_e$ be the same for both core-excited levels. The result of this fit  was not much different from the fit curve in Figure~\ref{fig:C2fit1}, i.e., some difference between the two core-excited potential curves seems to be a requirement for an improved fit.

In the fit, the rotational broadening was accounted for as discussed above. The resulting rotational temperature of $1095\pm137$~K corresponds  roughly to the expected temperature of the cesium vapor in our Cs-sputter ion source. Moreover, it was assumed that the same temperature also determines the populations of the $X\,^2\Sigma_g^+$ and $A\;^2\Pi_u$ vibrational levels. Accordingly, the relative populations $n_{kv}$ (Equation~\ref{eq:fit2}) of the $X\,^2\Sigma_g^+$ ($A\;^2\Pi_u$) $v=0$ and $v=1$ levels were 91\% (90\%) and 9\% (10\%), respectively. Within this approach, the initial populations of all higher vibrational levels were insignificant. As already mentioned above, larger contributions from higher vibrational levels would smear out the observed vibrational resonance structures and, thus, be at odds with the experimental observation.

In addition to the rotational broadening, also the lifetime broadening was obtained from the fit.  The present lifetime widths of 124(12)~meV and 154(21)~meV  (Table~\ref{tab:fit}) are  reasonably close to the more exact values for other carbon-containing small molecules such as, e.g.,  $99(2)$~meV for CO$_2$ \cite{Carroll2000} and $100(1)$~meV for C$_2$H$_2$ \cite{Hoshino2006}. In all fits, the instrumental width was kept fixed at its nominal value $\Delta E = 50$~meV.

\section*{Conclusions}

Using the photon-ion merged-beams technique, we have measured inner-shell photoabsorption of a molecular anion. The experimental approach is quite general and will be further exploited for investigating the interaction of energetic radiation with reactive molecular species. The favorable conditions with respect to photon flux and resolving power at beamline P04 of the PETRA\,III synchrotron light source enabled us to resolve individual vibrational transitions to core-excited C$_2^-$ molecular levels and to extract their potential parameters from a detailed Franck-Condon analysis, which consistently accounts for vibrational and rotational effects. This analysis revealed that the dicarbon anion shrinks substantially by 0.2~\AA\ upon $1\sigma$ core excitation. Similarly strong geometrical changes might also be expected for the photoexcitation of other anionic molecular systems.

The present results may aid the detection of C$_2^-$ in cold cosmic gas clouds where several hydrogenated C$_n$H$^-$ species have been identified owing to their large dipole moments \cite{Millar2017}. So far, the infrared-inactive C$_2^-$ anion has not been discovered in space. The presently measured C$_2^-$ double-detach\-ment cross section exhibits clear spectral signatures in the soft x-ray spectral range which can help to identify interstellar C$_2^-$ by upcoming high-resolution x-ray telescopes such as Athena \cite{Barret2020}. We also hope that the present work stimulates the further development of the theoretical tools (see, e.g., Huang et al. \cite{Huang2022}) for a more exact treatment of core-excited molecules.

\section*{Acknowledgements}

We acknowledge DESY (Hamburg, Germany), a member of the Helmholtz Association HGF, for the provision of experimental facilities. Parts of this research were carried out at PETRA\,III and we would like to thank Kai Bagschik,  Frank Scholz, J{\"o}rn Seltmann, and Moritz Hoesch for assistance in using beamline P04. We are grateful for support from Bundesministerium f{\"u}r Bildung und Forschung within the \lq\lq{}Verbundforschung\rq\rq\ funding scheme (Grant Nos.\ 05K19GU3, 05K19RF2, and 05K19RG3) and from Deutsche Forschungsgemeinschaft (DFG, Project No.\ 389115454). M.M.\ acknowledges support by DFG through project SFB925/A3.

\section*{Conflict of Interest}

The authors do not declare any conflict of interest.

\begin{shaded}
\noindent\textsf{\textbf{Keywords:} \keywords}
\end{shaded}

\setlength{\bibsep}{0.0cm}

%\bibliographystyle{Wiley-chemistry}
%\bibliography{/tex/AMPstefan.bib}

\end{document}